\documentclass[pteplogo, preprint]{ptephy_v2}
\usepackage{braket}
\usepackage{amsthm}
\usepackage{ascmac}
\usepackage{amssymb, amsmath}
\usepackage{bm}
\usepackage{physics}
\usepackage{graphicx}
\usepackage{color}
\usepackage{autobreak}
\usepackage{url}
\usepackage{times}
\usepackage[T1]{fontenc}
\usepackage[normalem]{ulem}
\usepackage{multirow}

\newcommand{\opt}{{\mathrm{opt}}}
\newcommand{\inner}[2]{\bm{#1}\cdot\bm{#2}} 
\newcommand{\mev}{{\mathrm{MeV}}}
\newcommand{\fm}{{\mathrm{fm}}}
\newcommand{\zi}{\,{\mathrm{i}}}
\preprintnumber{2408.14760}
\begin{document}
\title{Interrelation between $\bm{\bar{p}}$-Ca Atom Spectra and Nuclear Density Profiles}



\author{Kenta Yoshimura}
\author{Shunsuke Yasunaga}
\author{Daisuke Jido}


\affil{Department of Physics, School of Science, Tokyo Institute of Technology, Tokyo 152-8551, Japan}

\author{Junko Yamagata-Sekihara}
\affil{Department of Physics, Kyoto Sangyo University, Kyoto 603-8555, Japan}

\author{Satoru Hirenzaki}
\affil{Department of Physics, Nara Women's University, Nara 630-8506, Japan}

\date{\today}

\begin{abstract}
This work studies $\bar{p}$-Ca atom spectra in light of the strong shifts and level widths, using the optical model with several types of parametric coefficients.
The spectroscopic quantities are obtained as the eigenvalues of the Dirac equation, where the nuclear densities computed via nuclear Density Functional Theory and the effect of the anomalous magnetic moment are incorporated.
The results indicate that the systematical difference of strong shifts between $^{40}$Ca and $^{48}$Ca can never be elucidated by only the conventionally adopted isoscalar $b_0$ term, necessitating additional contributions from the isovector $b_1$ and p-wave $c_0$ terms.
Furthermore, it is found that both the strong shifts and the level widths exhibit significant dependence on the nuclear density profiles.
These findings demonstrate that, the detailed nuclear structures make a significant contribution to the observed spectrum, at least for the calcium isotopes.


\end{abstract}
\maketitle
\maketitle
\section{INTRODUCTION}

Hadronic atoms, which comprise negatively charged mesons or baryons substituting orbiting electrons, represent an intriguing topic within the nuclear-hadron physics, providing deeper insights into the properties of hadrons at finite density~\cite{batty1997a,friedman2007a}.
One of the most successful examples is the precise extraction of the in-medium quark condensate from the spectroscopy of deeply bound pionic atoms~\cite{nishi2023}.
Antiprotonic systems are also expected to contain significant features~\cite{batty1989, klempt2005,  doser2022}, such as $\bar{N}N$ interactions, QED testing and pair-annihilation processes~\cite{wycech2001,gaitanos2015,adamczyk2015, hrtankova2016, dai2017, aghai-khozani2018, hrtankova2018, germann2021,paul2021,haidenbauer2022},
where the atomic spectroscopy can offer novel perspectives.
Additionally, the rapid advancements in X-ray spectroscopic experiments enable us to deduce not only the intermediate hadron interactions but also nuclear structures~\cite{batty1989a, wycech1996c, lubinski1998, schmidt2003,jastrzebski2004, klos2004,klos2007,wycech2007, friedman2008,trzcinska2009, ficek2018,aumann2022} (and references therein), necessitating more precise investigations from theoretical points of view.

In the contexts of hadronic atoms, one interesting topic is the estimation of the {\it strong shifts} and {\it level widths}~\cite{wycech1993, gotta1999, gotta2004},
which originate from the strong force and characterize energy level shifts and particle absorptions.
These features have been described by the optical model with the aid of the linear density approximation~\cite{kisslinger1955, ericson1966,nishimura1976, batty1981, cote1982, batty1987, batty1995, friedman2004, friedman2005, friedman2015}.
The optical model integrates the strong force by introducing an additional potential $V_\opt = U - iW$, whose real part leads to the energy shift, while the imaginary part represents the absorption effect.
Within the linear density approximation it is assumed that each part of the optical potential depends on a linear combination of neutron and proton densities $\rho_n$ and $\rho_p$, with their coefficients parameterized to match experimental data.
Regarding these parameterization, while the relativistic mean-field (RMF) approach~\cite{reinhard1989,serot1992} has been adopted for investigating deeply bound states~\cite{burvenich2002,friedman2004, mishustin2005, larionov2008, larionov2010, gaitanos2011, hrtankova2016}, for level shifts and widths global fittings across wide-range measured atoms have been prevailing~\cite{friedman2013, friedman2014, friedman2019, friedman2019a}.
For antiprotonic atoms this global fitting protocol with the recent PS209 experiments has been successfully conducted~\cite{batty1995, friedman2004, friedman2005}, concluding that contributions from neutrons and protons should be symmetric.
This indicates that density dependence of the optical potential can be described by solely the isoscalar density $\rho_0 = \rho_n + \rho_p$.
On the other hand, the experiments have demonstrated significant differences in both shifts and widths among isotope nuclei~\cite{trzcinska2001}, some of which cannot be explained by the potentials possessing only the isoscalar term, as pointed out in Ref.~\cite{hartmann2001a, klos2007}. 
In order to quantitatively investigate such individual spectral behaviors, it is necessary not only to determine a global potential by fitting but also to construct and examine a tailored model for each nucleus. 
Furthermore, in this process, there is room to consider adding new terms to the current optical model.


The present paper focuses on Calcium isotopes, especially $^{40}$Ca and $^{48}$Ca.
Although both are known as double magic nuclei, their properties remain not fully understood.
Firstly, whereas in general charge radii swell up as the neutron number increases, the counterparts of these two nuclei are almost identical, exhibiting a ``kink'' structure between atomic number 40 and 48~\cite{naito2023a}.
Additionally, there are also discrepancies related to the density distribution, as a recent experimental result has shown that the density profile of $^{40}$Ca is significantly different from any theoretical prediction~\cite{zenihiro2018, zenihiro2021, sagawa2022}.
Investigations of $\bar{p}$-atom spectroscopy could provide a novel perspective on this perplexing situation.
Our work suggests a theoretical and comprehensive approach to $\bar{p}$-atom studies as a tool for the extraction of the nuclear properties.
The points which this study focuses on are as follows.
In the first place, we examine the role of other terms in the optical potential than the conventional isoscalar term.
It is known that there is a systematical difference by a factor of several between the strong shifts of $^{40}$Ca and $^{48}$Ca, although the optical model with the prevailing parameter $b_0 = (1.5+2.5i)\,\fm$ predicts around 3 eV for both of them~\cite{hartmann2001a}.
This justifies that, at least for Calcium isotopes, additional contributions such as the isovector $b_1$ term and/or the p-wave term should be considered in the optical potential.
In the second place, more realistic density profiles are adopted in this work.
Many of the precedent researches have utilized simple 2-parameter or 3-parameter Fermi (2pF, 3pF) models with phenomenological parameters such as ``skin'' or ``halo'' types~\cite{garcia-recio1992, schmidt1998, krasznahorkay1999, trzcinska2001a, clark2003, schmidt2003, jastrzebski2004}.
This is because, while the proton densities have been precisely obtained through electron scattering experiments~\cite{devries1987}, 
it is difficult to get neutron densities directly.
For the meanwhile, in the field of nuclear physics, Density Functional Theory (DFT) calculations integrating the Skyrme effective interactions~\cite{skyrme1958,vautherin1972} have been successfully developed for the description and prediction of nuclear structures~\cite{colo2020}.
It is worthwhile investigating how the detailed density profiles impact the spectroscopic observables.
In the third place, we consider the effect of the spin-orbit splittings.
Within many conventional studies of antiprotonic atoms related to level shifts and widths, calculations have been performed using the Klein-Goldon equation, in which it has been presumed that the spin-orbit splittings are too small to distinguish experimentally.
However, due to the large anomalous magnetic moment of antiprotons, a planned high-resolution X-ray spectroscopic experiment with a state-of-the-art quantum sensor~\cite{higuchi2025} could separate the split energy levels.
A proper quantum mechanical treatment of spin $1/2$ fermions is the Dirac equation, whereby the spin-orbit interaction is inherently taken into account.

In this context, we develop a calculation framework based on the Dirac equation, integrating the optical model and the effect of the amonalous magnetic moment.
In the calculations, we assume the spherical symmetry taking into account the sphericity of both $^{40}$Ca and $^{48}$Ca, leading to lower computational cost.
For the density profiles, we calculate densities using not only the simple 3pF model, but also the DFT models with several parameter sets.
The coefficients for the optical potential have been essentially prepared referring to some of the precedent researches, where they have dealt with the global fitting in alignment with the experimental data~\cite{batty1995, batty1997a}.
In the meanwhile, we also depart from them to explore new combinations of parameters with considering the reproducibility of the experimental results of the Ca isotopes.

Because the lowest state dominantly observed in X-ray spectroscopy is the $(n,l)=(5,4)$ state for Ca isotopes~\cite{hartmann2001a}, in this paper we focus on the properties of $n=5$ and $n=6$ orbits.
By comparing the results with various density profiles and optical potential models, we investigate how nuclear structures and potential shapes influence spectroscopic quantities such as strong shifts and level widths.
It must be emphasized that this study does not propose a new set of global parameters intended to replace the existing globally fitted potential. 
One should note that the present discussion is strictly concerned with the case of calcium isotopes.

The arrangement of this article is as follows.
Section II explains the formalism and methodologies used in this work.
Section III presents and discusses the calculation results.
Section IV is dedicated to further discussions of the calculation results.
Section V summarizes the work and provides the future prospects.

\section{FORMALISM}
This section formulates the equation of motion (EoM) for the $\bar{p}$-Ca atomic systems. 
The EoM is expressed as Dirac equation, incorporating both the electromagnetic potential and the optical potential representing the effect of the strong interaction.
For the optical potential, it is necessary to assume the shape of the potentials with its parametric coefficients and nuclear properties.
In this work we consider three models referring to the precedent research.
While physical quantities regarding the nuclear structures are adopted from the experimental data, the density profiles are given by the 3pF model and nuclear DFT.
In our calculations the EoM is written down into the matrix representation in the coordinate space, and the energy eigenvalues are obtained by diagonalization methods.

\subsection{Dirac equation}
When the scalar potential $S(\bm{r})$ and $0$-th component of the vector potential $V_0(\bm{r})$ are present, the Dirac Equation for 4-spinor $\psi(\bm{r})$ at a fixed energy $E$ is written as
\begin{equation}
    \left(-i\inner{\alpha}{\nabla}+\beta (m + S) + V_0 - \frac{\zi \kappa}{2m}\dv{V_0}{r}\bm{\gamma}\cdot\hat{\bm{r}}\right)\psi = E\psi,
\end{equation}
where the Pauli term with the anomalous magnetic moment $\kappa = 1.79284734$ is taken into account~\cite{smorra2017}.
In the standard representation of the Dirac matrices, we have $\bm{\alpha} = \mqty(0&\bm\sigma\\\bm\sigma&0)$, $\beta = \mqty(1&0\\0&-1)$, and $\bm\gamma = \mqty(0&\bm\sigma\\-\bm\sigma&0)$.
When both of the potential $S$ and $V_0$ are spherically symmetric, the wave function $\psi$ can be labeled by total angular momentum $j$, orbital angular momentum $l$ and azimuthal angular momentum $m$. Here, from the composition of orbital angular momentum and spin-1/2, $l$ takes the value of either $j+1/2$ or $j-1/2$ for given $j$. Thus, we use the shorthand notation $(\pm)$ for the case $l=j\pm1/2$, respectively. The wave function $\psi$ can be separated into radial and spherical variables as follows:
\begin{equation}
    \psi_{jm}^{(\pm)}(\bm{r})=\mqty(i\frac{G_{j}^{(\pm)}(r)}{r}\mathcal{Y}_{jm}^{(\pm)}(\theta,\phi)\\\frac{F_j^{(\pm)}(r)}{r}(\inner{\sigma}{\hat{r}})\mathcal{Y}_{jm}^{(\pm)}(\theta,\phi)),
\end{equation}
where $\mathcal{Y}_{jm}^{(\pm)}$ is the spinor spherical harmonics defined as
\begin{equation}
    \mathcal{Y}_{jm}^{(\pm)}=\mqty(\braket{(\pm),m-\frac{1}{2},\frac{1}{2},+\frac{1}{2}}{j,m}Y_{(\pm),m-\frac{1}{2}}\\\braket{(\pm),m+\frac{1}{2},\frac{1}{2},-\frac{1}{2}}{j,m}Y_{(\pm),m+\frac{1}{2}}).
\end{equation}
Under this variable separation, the EoM for the radial directions can be written as 
\begin{equation}
    \begin{aligned}
        (E &- m - S(r) - V_0(r))G_j^{(\pm)}(r) \\
        &= -\dv{F_j^{(\pm)}(r)}{r}\pm\left(j+\frac{1}{2}\right)\frac{F_j^{(\pm)}(r)}{r} - \frac{\kappa}{2m}\dv{V_0}{r}F_j^{(\pm)}(r),\\
        (E &+ m + S(r) - V_0(r))F_j^{(\pm)}(r) \\
        &= \dv{G_j^{(\pm)}(r)}{r}\pm\left(j+\frac{1}{2}\right)\frac{G_j^{(\pm)}(r)}{r} - \frac{\kappa}{2m}\dv{V_0}{r}G_j^{(\pm)}(r).\label{Eq:RadialDirac}
    \end{aligned}
\end{equation}

Given the two-body systems with masses $m_A$ and $m_B$, and the interactions exclusively depending on the relative distance between them, the EoM can be divided into the center of the mass part and the relative coordinate part.
For the latter part, the form of the EoM is the same with the original Dirac equation (\ref{Eq:RadialDirac}), except that the mass term is replaced by the {\it reduced mass}, written as
\begin{equation}
    \mu = \frac{m_Am_B}{m_A+m_B}.
\end{equation}

In the following formulations, we implicitly assume that the spherical symmetry holds in interested systems, and the coordinate $r$ always stands for the relative distance between the center of mass of two particles, namely the Ca nuclei and antiprotons.

\subsection{Coulomb Potential}
In the $\bar{p}$-Ca atomic systems, the 0-th component of the vector potential $V_0$ contains the electromagnetic potential $-eA_0$ in the Coulomb gauge:
\begin{equation}
    -eA_0(r) = -\alpha\int\dd^3\bm{r}' \frac{\rho_c(r')}{\abs{\bm{r}-\bm{r}^\prime}},\label{eq:Coulomb}
\end{equation}
where $\alpha = e^2 / 4\pi\epsilon_0 = 1/137.036$ is the fine structure constant and $\rho_c$ is the charge density which is given later.
With the integral for elevation and azimuth angle components performed, it turns into
\begin{equation}
    -eA_0(r) = -\frac{2\pi\alpha}{r}\int_0^\infty \dd r'r'\rho_c(r')(r+r' - |r-r'|).
\end{equation}
In our framework, we consider the effect of the vacuum polarization up to the first-leading $Z\alpha^2$ order.
Let us write this contribution as $V_{\mathrm{VP}}$ and take the method proposed by Refs.~\cite{nieves1993, ikeno2015a}:
\begin{equation}
    V_{\mathrm{VP}}(r) = -\frac{2}{3}\frac{\alpha^2}{m_er} \int_0^\infty \dd r^\prime r^\prime\rho_c(r^\prime)\qty[f(|r-r^\prime|) - f(r+r^\prime)],
\end{equation}
with
\begin{equation}
    f(r) = \int_1^\infty dt e^{-2m_e rt} \qty(\frac{1}{t^3} + \frac{1}{2t^5}) \sqrt{t^2-1},\label{eq:vp-f}
\end{equation}
which corresponds to $K_0(x)$ in Ref.~\cite{fullerton1976} and where $m_e$ is the electron mass.
Finally the vector potential can be written as
\begin{equation}
    V_0(r) = -eA_0(r) + V_\mathrm{VP}(r).
\end{equation}
Although, in Ref.~\cite{borie1983}, it has been pointed out that the higher order of the vacuum polarization may change the binding energies by several percent, in the present paper this effect is not considered.
This is because we calculate the energy shifts by the strong interaction which are given by the difference between the binding energies with and without the optical potential and thus this difference should not be affected largely by the detail shapes of Coulomb potential.
This is also the case for the level widths because they mainly depend on the imaginary part of the optical potential.

\subsection{Optical Potential and density folding}
In our framework, the optical potential is contained in the scalar potential $S(r)$.
As explained previously, within the framework of the linear density approximation, the optical potential can be represented as a linear combination of the nuclear (neutron and proton) densities $\rho_n(r)$ and $\rho_p(r)$:
\begin{equation}
    S(r) = [B_n \rho_n(r) + B_p\rho_p(r)] + \grad[C_n\rho_n(r) + C_p\rho_p(r)]\cdot\grad.
\end{equation}
A generic form of the optical potential has the following form:
\begin{equation}
    \begin{aligned}
        S(r) =&-\frac{4\pi}{2\mu}\qty(1+\frac{\mu}{M_N}) \qty[b_0\rho_0(r) + b_1\rho_1(r)]\\
        &+\frac{4\pi}{2\mu}\qty(1+\frac{\mu}{M_N})^{-1}\grad\qty[c_0\rho_0(r) + c_1\rho_1(r)]\cdot\grad,
    \end{aligned}\label{eq:opticalmodel}
\end{equation}
with the isoscalar $\rho_0$ and isovector density $\rho_1$ written as
\begin{eqnarray}
    \rho_0(r) &=& \rho_n(r) + \rho_p(r),\\
    \rho_1(r) &=& \rho_n(r) - \rho_p(r),
\end{eqnarray}
and the averaged nucleon mass $M_N=938.918\,\mev$.
The terms in the first and second braces represent the s-wave and p-wave components of the relative partial wave, respectively, whose detailed derivation is explained in Ref.~\cite{ericson1966,kisslinger1955}.
In practical usage, it is often the case that the particle densities in Eq.~\eqref{eq:opticalmodel} are replaced by their ``folded'' densities~\cite{friedman2005} for taking into consideration the finite size effect of particles.
In this work, we employ a simple Gaussian form folding scheme:
\begin{equation}
    \rho^F(r) = \int \dd\bm{r}^\prime \rho(\bm{r}^\prime) \frac{1}{\pi^{3/2}\beta^3} e^{-(\bm{r}-\bm{r}^\prime)^2 / \beta^2},\label{eq:fold}
\end{equation}
referring to the precedent researches.
In this formula, $\beta$ represents the size of nucleons and we adopt $\beta=0.85\,\fm$ which is the same with Ref.~\cite{friedman2005}.
These folded densities for protons can be identified with the charge density in Eq.~\eqref{eq:Coulomb}.


The coefficients $b_0$, $b_1$, $c_0$, $c_1$ could be connected to the scattering lengths and volumes of $\bar{N}N$ scatterings by means of the $t\rho$ approximation, and their values have been calculated for example by Paris potential~\cite{cote1982,pignone1994,el-bennich1999, el-bennich2009, friedman2015}.
However, it is known that potentials derived from the scattering lengths do not reproduce the experimental results for $\bar{p}$ atom spectra.
This is because nuclei are in general many nucleon systems where they interact each other and form bound states, and for precise predictions of their properties many-body effects should be taken into account.
Therefore, it is common practice to treat these coefficients as free parameters which should be fitted to the experimental data of the spectroscopy, rather than the scattering cross section.
In the sense that the many-body effects are ``integrated'' in the linear density terms, their values can be magnitude of orders different from their original scattering lengths.
For the determination of parameters, global fittings referring to the latest PS209 experiments \cite{trzcinska2001,friedman2004} have been prevailing, which reach the consensus that neither the isovector part nor p-wave component contribute significantly.
In this paper, however, we employ the optical potentials with these expelled components, because we are mainly interested in the systematical relationships between the optical models and the physical quantities, rather than in a quantitatively precise prediction of the $\bar{p}$-Ca spectroscopy.
In this sense this treatment of optical potentials should be justified,
and for this purpose we prepare a few characteristic parameter sets, which are exhibited in Table ~\ref{tab:opticalparameters}.
All parameter sets have the same value for the $b_0$ terms, which is the same with the results of recent Friedman's global fitting~\cite{friedman2004, friedman2005}, and the difference is whether the isovector or p-wave term is included or not.
Regarding to the $c_0$ component, we set a finite value for only the imaginary part,
which is inferred from Wycech's best fit~\cite{wycech2007}.
The isovector $b_1$ part is referring to the results of global fitting~\cite{batty1995b}, but its value is adjusted considering the effect of density folding.
In Ref.~\cite{batty1995b} the fitting result yields $b_1 = -12.4 + 4.2i\,\fm$, while in this work $b_1 = -8.0 + 1.7i\,\fm$ is adopted.
They should enable us to extract the role of each term integrated in the optical potential.

\begin{table}[tbp]
    \centering
    \caption{The optical potential parameter sets used in this work.}
    \begin{tabular}{cccc}
         & $b_0(\fm)$ & $b_1(\fm)$ & $c_0(\fm^3)$  \\
        \hline\hline
        Type I & $1.3+1.9\zi$ & -- & --\\
        Type II & $1.3+1.9\zi$ & $-8.0+1.7\zi$ & --  \\
        Type III & $1.3+1.9\zi$ & -- & $0.0 + 1.2\zi$ \\ 
        \hline
    \end{tabular}
    \label{tab:opticalparameters}
\end{table}

\begin{table}[tbp]
    \centering
    \caption{Parameters for the 3pF model in $^{40}$Ca and $^{48}$Ca, with simplified neutron skin thickness. From left to right, the mass of nuclei $M_{\mathrm{Ca}}$, the proton mean radius $R_p$, diffuseness $a$, $w$, and the neutron mean radius $R_n$ are exhibited.}
    \begin{tabular}{cc|ccccc}
        && $M_{\mathrm{Ca}}$(MeV) & $R_p$(fm) & $a$(fm) & $w$ & $R_n$(fm)\\
         \hline\hline$^{40}$Ca & 3pF & $37214.698$ & $3.766$ & $0.586$ & $-0.161$ & $3.766$ \\
         $^{48}$Ca & 3pF-2.0 & $44657.272$ & $3.7369$ & $0.5245$ & $-0.03$ & $3.9369$\\
         & 3pF-2.5 &&&&& $3.9869$\\
         & 3pF-3.0 &&&&& $4.0369$\\
         \hline
    \end{tabular}
    \label{tab:3pFparameters}
\end{table}

\begin{table}[htbp]
    \centering
    \caption{A comparison table of proton and neutron mean radii for both $^{40}$Ca and $^{48}$Ca, with densities calculated by the 3pF model with empirical parameters, or by the DFT model with several profiles. All the quantities are in a unit of fm.}
    \begin{tabular}{cccccccc}
        & \multicolumn{3}{c}{$^{40}$Ca} && \multicolumn{3}{c}{$^{48}$Ca}\\
         \cline{2-4}\cline{6-8} & $r_n$ &  $r_p$ & $\Delta r_{np}$ &&  $r_n$ &  $r_p$ & $\Delta r_{np}$\\
        \hline\hline 3pF-2.0 & 3.482  & 3.482  & 0.00     && 3.600  & 3.469  & 0.130  \\
    3pF-2.5 & 3.482  & 3.482  & 0.00     && 3.633  & 3.469  & 0.163  \\
    3pF-3.0 & 3.482  & 3.482  & 0.00     && 3.665  & 3.469  & 0.196  \\
    \hline SLy4   & 3.531  & 3.576  & $-0.045$ & & 3.755  & 3.608  & 0.147  \\
    SLy5   & 3.526  & 3.572  & $-0.046$  && 3.760  & 3.606  & 0.154  \\
    SkM*   & 3.535  & 3.582  & $-0.047$  && 3.749  & 3.600  & 0.149  \\
    SAMi   & 3.502  & 3.547  & $-0.045$  && 3.762  & 3.595  & 0.168  \\
    SGII   & 3.507  & 3.555  & $-0.048$  && 3.731  & 3.590  & 0.141  \\
    UNEDF0 & 3.541  & 3.581  & $-0.040$  && 3.794  & 3.592  & 0.201  \\
    UNEDF1 & 3.521  & 3.555  & $-0.034$ & & 3.767  & 3.590  & 0.177  \\
    UNEDF2 & 3.510  & 3.545  & $-0.035$ & & 3.746  & 3.585  & 0.160  \\
    HFB9   & 3.534  & 3.578  & $-0.045$  && 3.754  & 3.603  & 0.151  \\ \hline
    \end{tabular}
    \label{tab:Skyrmeparameters}
\end{table}

\begin{figure}[ht]
    \centering
    \includegraphics[width=\columnwidth]{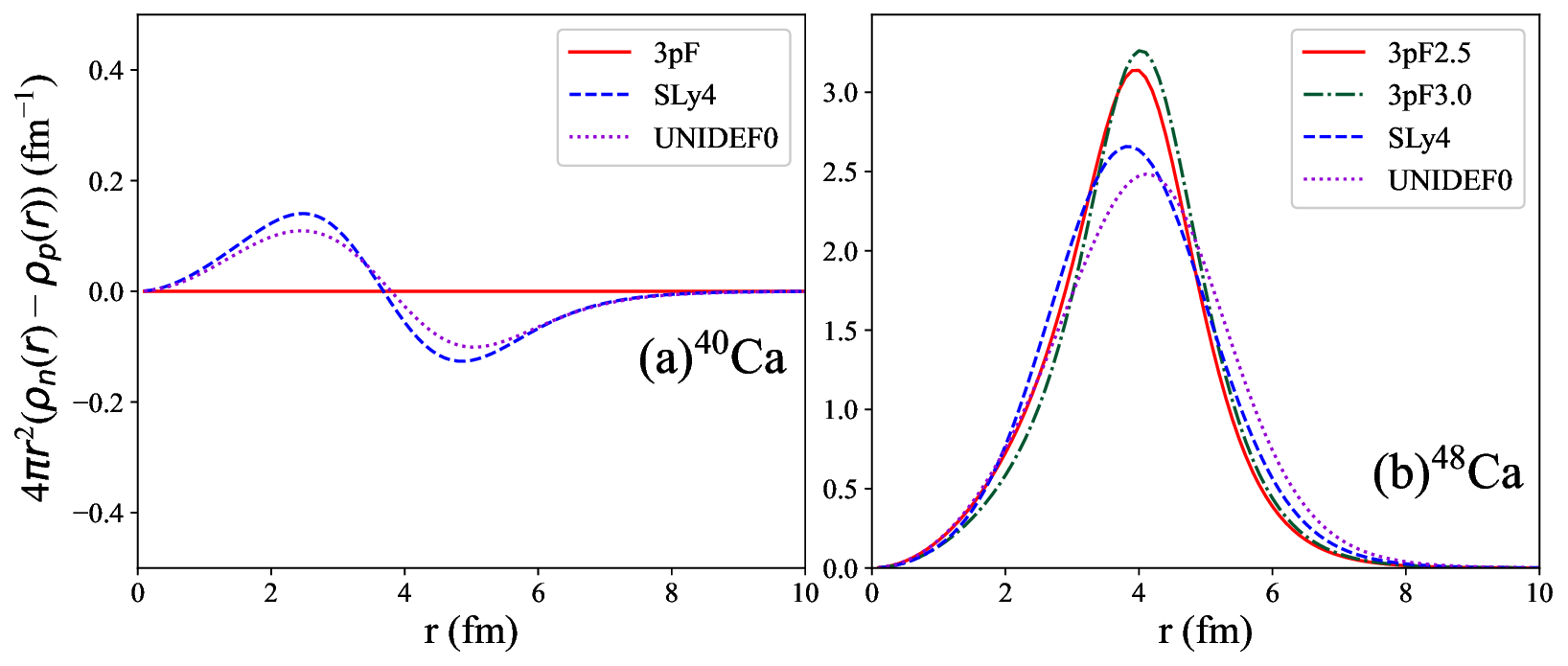}
    \caption{The isovector density $\rho_1$ as a function of the radial coordinate, with respect to several density profiles. In the left panel, distribution of $^{40}$Ca is exhibited, and the right shows that of $^{48}$Ca.}
    \label{fig:Skyrme_denscomp}
\end{figure}

\begin{figure}[ht]
    \centering
    \includegraphics[width=\columnwidth]{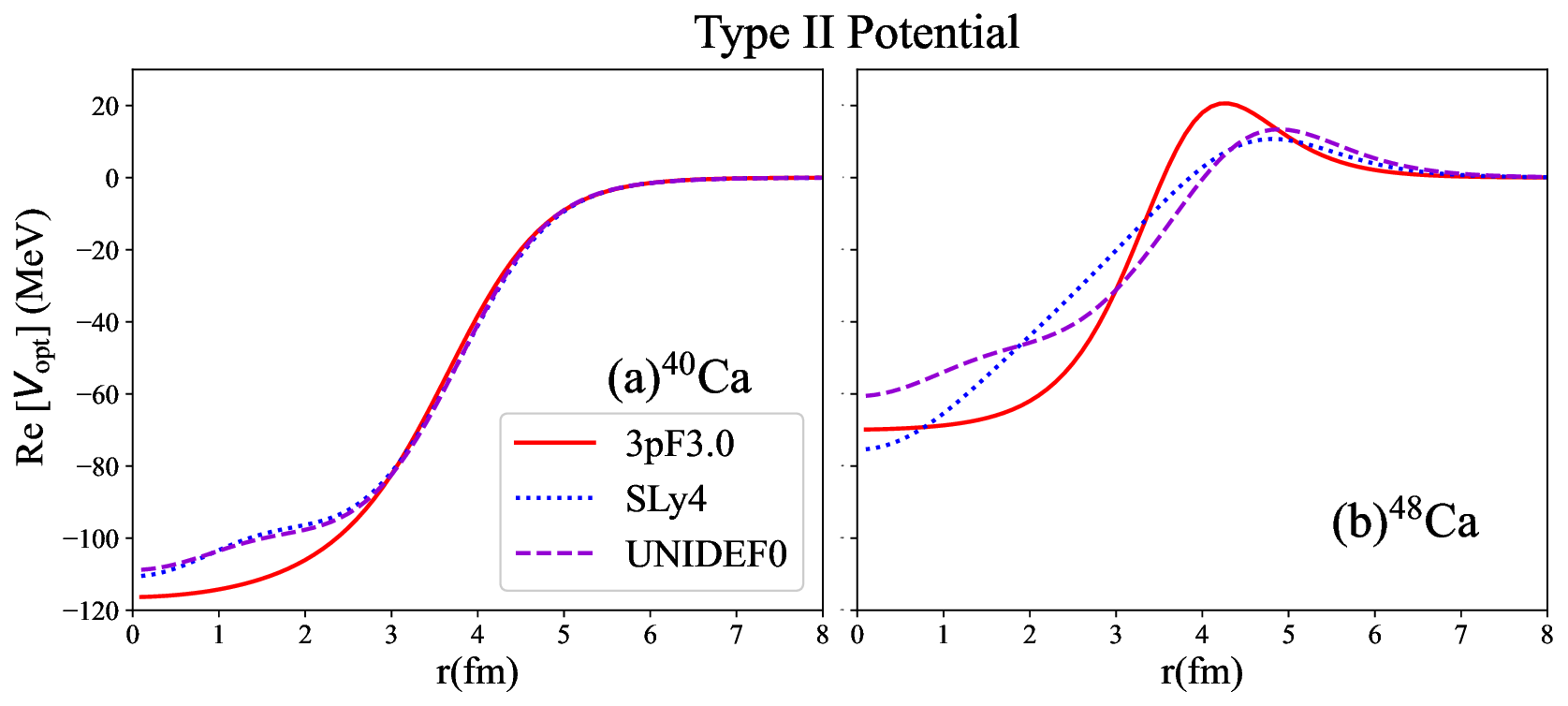}
    \caption{The real part of the optical potential with Type II parameters as a function of the radial coordinate, for several representative of the nuclear profiles. In the left panel, distribution of $^{40}$Ca is exhibited, and the right shows the counterpart of $^{48}$Ca.}
    \label{fig:optical_typeII}
\end{figure}

\subsection{Nuclear Properties}
To perform the calculations and extract the quantitative observables from the above formalism, we need the information on the nuclear properties.
Among them, the masses of nucleons $M$ and each nucleus $M_{\mathrm{Ca}}$ are experimentally evaluated and available in Ref.~\cite{zotero-626}.
On the other hand, there are several different models for nuclear densities, especially neutron densities.
Conventionally used has been the 3pF model where the particle density is simply represented by 
\begin{equation}
    \rho_{c,q}(r) = \qty(1+w\frac{r^2}{R_q^2})\frac{\rho_0}{1+\exp((r-R_q)/a)},
\end{equation}
with particle species $q=n,p$ and parameters $w$, $R_q$ and $a$.
The parameter $R_q$ stands for the nuclear radius and should be evaluated for neutrons and protons, respectively,
while $\rho_0$ is the normalization factor which should be determined consistently with the total particle numbers through
\begin{equation}
    \int 4\pi r^2\rho_{c,q}(r) \dd r = N_q.
\end{equation}
Whereas the proton charge radius $R_p$ is well known thanks to electron scattering experiments, there is a little information on neutron radius $R_n$, which necessitates to assume the phenomenological function form.
For $^{40}$Ca, a symmetric nucleus, we can assume that the neutron and proton densities are identical as a good approximation.
However, for neutron-rich $^{48}$Ca, such an assumption may not be considered realistic, due to the existence of the neutron ``skin'' at the nuclear surface.
Regarding the neutron skin thickness, there is still a room for discussion, so that in this work, we prepare three models for $R_n$, which we call 3pF-2.0, 3pF-2.5, 3pF-3.0, respectively. 
In Table \ref{tab:3pFparameters}, the parameters for $^{40}$Ca and $^{48}$Ca are exhibited.
The masses of nuclei are adopted from~\cite{zotero-626}, and the others except the neutron radii $R_n$ are referring to~\cite{devries1987}.
As explained above, for $^{40}$Ca we set the neutron radius as $R_n = R_p$, and for 3pF-X.X of $^{48}$Ca, we set $10(R_n - R_p) =$ X.X.

In addition to the 3pF model, we employ more realistic density distributions obtained through the nuclear DFT calculations~\cite{vautherin1972}.
Within the DFT framework, the ground state is calculated according to the variational principle of the energy density functionals (EDFs), with several parameter sets.
There exist a lot of parameter sets in accordance with their dedicated experimental database, and we adopt several representative parameter sets of the Skyrme-type EDF~\cite{vautherin1972}; SLy4~\cite{chabanat1998}, SLy5~\cite{chabanat1998}, SkM*~\cite{bartel1982}, SAMi~\cite{roca-maza2012}, SGII~\cite{vangiai1981a}, UNEDF0~\cite{kortelainen2010a}, UNEDF1~\cite{kortelainen2012}, UNEDF2~\cite{kortelainen2014}, HFB9~\cite{goriely2005}.
We prepare density profiles generated with these parameters based on the spherical Hartree-Fock-Bogoliubov calculations~\cite{dobaczewski1984},
where the volume-type pairing interaction is used whose strengths are given in Ref.~\cite{naito2023a}.
While the densities inferred from the 3pF model stand for the charge density, ones calculated from the Skyrme model is in general point particle densities, which need the folding process already indicated in Eq.~\eqref{eq:fold} for the optical potential.
The root-mean-square (RMS) radii of neutron and proton densities after the folding process are summarized in Table~\ref{tab:Skyrmeparameters}.

Figure \ref{fig:Skyrme_denscomp} shows the isovector densities $\rho_1 = \rho_n - \rho_p$ for $^{40}$Ca and $^{48}$Ca in the left and right panels, respectively.
For $^{40}$Ca, the density of the 3pF model is always zero, which is natural because all the parameters are same,
while the counterparts of the Skyrme model go to negative at the surface.
For $^{48}$Ca, although isovector densities are always positive for all density profiles, the position and height of peaks differ,
which should lead to the significant variety of the Type II optical potential.
Figure \ref{fig:optical_typeII} demonstrates the shape of the real part of the optical potential with the Type II potential, for several representative density profiles.
From this figure, one can find three characteristic behaviours.
In the first place, the optical potentials of $^{40}$Ca and $^{48}$Ca are significantly different.
This difference can be attributed to the effects of the isovector term, because the isovector density of $^{40}$Ca is much smaller than $^{48}$Ca and does not largely alter the potential shape.
In the second place, for $^{40}$Ca, density profiles in case of DFT make the potential shallower in the center region.
This point itself seems to affect nor the strong shifts neither level widths, because the $\bar{p}$-atom spectra are considered to depend on only the nuclear periphery structures.
However, the depth of the potential may alter the level structures of deeply bound states, which could lead to the shift of the atomic states through the level repulsion.
In the third place, for $^{48}$Ca, their density distributions are totally different.
In the periphery region, it is found that the potentials change to repulsive, due to the negative value of $\Re{b_1}$.
The magnitude of these repulsive regions is significantly different among the density profiles, which should impact the spectroscopic observables quantitatively.
\subsection{Computation Method}
In the present work, the energy eigenvalues $E$, generally complex numbers, of the Dirac equation \eqref{Eq:RadialDirac} are calculated via matrix diagonalizations of the Hamiltonian matrix for the radial wave functions $F(r)$ and $G(r)$.
For the bound states, the obtained eigenenergies can be connected to the binding energy $B$ and the absorption width $\Gamma$, by using the following relation:
\begin{equation}
    E = \mu - B - i\frac{\Gamma}{2}.
\end{equation}
Because both the reduced mass $\mu$ and binding energy $B$ are always real, the absorption width can be easily obtained from the imaginary part of the energy eigenvalues.
On the other hand, the strong shifts cannot be obtained directly,
as they represent the difference in binding energies calculated with and without the optical potential.
Let us define the binding energy with the optical potential as $B_{\mathrm{w/opt}}$ and without the optical potential $B_{\mathrm{no.opt}}$.
The strong shift $\varepsilon$ is calculated by
\begin{equation}
    \varepsilon = B_{\mathrm{no.opt}} - B_{\mathrm{w/opt}}.
\end{equation}

For constructing the Hamiltonian matrix, we employ the finite difference method (FDM), in which the radial coordinate space is discretized into $N_r$ grid points with spacing $\Delta r$.
The size of the cell $R_{\mathrm{max}}$ is determined by $R_{\mathrm{max}} = N_r\Delta r$.
Because atomic orbits extend far outside nuclei, a small cell size may conflict with the boundary condition $\eval{F(r)}_{r=\infty}=\eval{G(r)}_{r=\infty}=0$, leading to fluctuations in both the wave functions and energy eigenvalues.
We verify that the setting $N_r=1200$ and $\Delta r=0.2\,\fm$ provides good convergence for any bound state examined.
\section{RESULTS}
In this section, we show the calculation results.
Our discussing issues are mainly three-folds;
1.~what optical model can reproduce the systematical differences of strong shifts and level widths between $^{40}$Ca and $^{48}$Ca,
2.~how the density profiles impact the $\bar{p}$-Ca atom spectroscopy,
3.~whether the spin-orbit splitting can be distinguished experimentally, under the effect of the anomalous magnetic moment. 
All these issues will be thoroughly reviewed separately in this section.
While the tables of the calculation results are arranged in align with each dedicated subsection, the full set of results can be seen in Appendix \ref{Sec:FullResult}.
\subsection{Pure Electromagnetic States}
First of all, we show the calculation results of the binding energies in Table \ref{tab:bindingenergies} without the optical potential implemented.
For comparison, we juxtapose the calculated binding energies by means of the exact solution from the point Coulomb potential, which is represented by
\begin{equation}
    \begin{aligned}
        E^{\mathrm{PC}}_{nj} =& \mu \qty[1 + \frac{Z^2\alpha^2}{\qty{n-j-\frac{1}{2} + \sqrt{\qty(j+\frac{1}{2})^2 - Z^2\alpha^2}}^2}]^{-1/2}
    \end{aligned}
\end{equation}
with the principle quantum number $n$ and total angular momentum $j$.
From Table ~\ref{tab:bindingenergies}, it is found that the calculated binidng energies in our framework are different from the point-Coulomb results by around zero point several percent, which can be attributed to the effects of the nuclear size, vacuum polarization and the anomalous magnetic moment.
The latter is because the calculated energies of both spin states are shifting in a direction that enhances the magnitude of the spin-orbit splitting.
This point will be discussed to be detailed in the later section.

\begin{table}[htbp]
    \centering
    \caption{The calculated binding energies of $(n,l)=(5.4)$ and $(6,5)$ states for $^{40}$Ca and $^{48}$Ca, by means of our suggesting framework, as well as the point-Coulomb potential. The optical potential is not included, and all energies are in a unit of keV.}
    \begin{tabular*}{0.5\columnwidth}{@{\extracolsep{\fill}}ccccc}
         &  \multicolumn{2}{c}{$^{40}$Ca} & \multicolumn{2}{c}{$^{48}$Ca}\\
        \cline{2-3}\cline{4-5} & calc. & PC & calc. & PC\\
        \hline\hline 
        5g$_{9/2}$ & 389.852 & 389.967 & 391.443 & 391.572\\
        5g$_{7/2}$ & 390.153 & 390.050 & 391.826 & 391.656\\
        6h$_{11/2}$ & 270.697 & 270.793 & 271.858 & 271.907\\
        6h$_{9/2}$ & 270.844 & 270.825 & 272.005 & 271.940\\
        \hline
    \end{tabular*}
    \label{tab:bindingenergies}
\end{table}

\subsection{Optical Model Comparison}

Table 5 lists, for the $g_{9/2}$ and $h_{11/2}$ orbitals respectively, the calculated strong shifts and level widths obtained with the Type I, Type II, and Type III optical potentials.  For comparison, results are shown for several representative density profiles, namely 3pF3.0, SLy4, UNEDF0, and HFB9, and the experimental values are provided at the bottom.
From these data the following points become clear for each optical-potential type.
In the first place, in Type I the shifts and widths of $^{40}$Ca and the width of $^{48}$Ca are reproduced reasonably well, while the shift of $^{48}$Ca deviates greatly from the experimental values for every density profile.  
This result is consistent with Ref.~[50], where fitting the neutron skin yielded good agreement for the widths while failed to reproduce the shifts.  
At the same time the result implies that, at least for calcium, the $b_0$ parameter obtained from global fitting is plainly insufficient and that an additional contribution is needed.
In the second place, for Type II the shifts of $^{48}$Ca is reproduced better than with the other two parameter sets.  By contrast, the shift of $^{40}$Ca have opposite signs between the 3pF profile and the others, which may be, as will be discussed later, attributed to the changes in the nuclear states induced by the complicated potential shape.
In addition, the width of $^{48}$Ca tends to be smaller than in Type I for every density profile due to the shallower potential bulk.
In the third place, in Type III the shifts of both $^{40}$Ca and $^{48}$Ca increase relative to the other types and approach the experimental values for all density profiles, whereas the widths grow considerably larger and diverge from experiment. 

Overall, the experimental shifts and widths of $^{40}$Ca and $^{48}$Ca cannot be reproduced satisfactorily with only the $b_0$ parameter obtained from conventional global fitting, which indicates that additional contributions that were previously considered negligible should be necessarily introduced.  Although the Type II and Type III optical models, which incorporate isovector and p-wave terms, on the one hand improve agreement with experiment in certain respects compared with Type I, whereas they fall short in others, so their parameter choices require more careful scrutiny.

\begin{table}[t]
    \centering
    \caption{The strong shift for $n=5$ and level width for $n=6$ in a unit of eV, with respect to optical parameter sets. The second to the third column, the result of $^{40}$Ca is provided and the fourth to fifth column is dedicated to the counterpart of $^{48}$Ca with the 3pF, SLy4, UNEDF0, HFB9 density profiles. At the bottom line the experimental result~\cite{hartmann2001a} is shown.}
    \begin{tabular*}{0.6\columnwidth}{@{\extracolsep{\fill}}ccccc}
         & \multicolumn{2}{c}{$^{40}$Ca} & \multicolumn{2}{c}{$^{48}$Ca} \\
         \cline{2-3}\cline{4-5} 3pF & $\epsilon_{g9/2}$ & $\Gamma_{h11/2}$ & $\epsilon_{g9/2}$ & $\Gamma_{h11/2}$\\
         \hline Type I & $2.32$ & $0.136$  & $3.67$ & $0.109$\\
         Type II & $2.32$ & $0.136$ & $14.5$ & $0.093$\\
         Type III & $7.73$ & $0.237$ & $10.7$ & $0.198$\\
         \hline\\
         \cline{2-3}\cline{4-5}SLy4& $\epsilon_{g9/2}$ & $\Gamma_{h11/2}$ & $\epsilon_{g9/2}$ & $\Gamma_{h11/2}$\\
        \hline Type I & $0.940$ & $0.048$  & $2.32$ & $0.076$\\
         Type II & $-2.56$ & $0.047$ & $20.3$ & $0.071$\\
         Type III & $8.26$ & $0.129$ & $12.9$ & $0.191$\\
         \hline\\
         \cline{2-3}\cline{4-5} UNEDF0& $\epsilon_{g9/2}$ & $\Gamma_{h11/2}$ & $\epsilon_{g9/2}$ & $\Gamma_{h11/2}$\\
        \hline Type I & $0.783$ & $0.051$  & $2.00$ & $0.087$\\
         Type II & $-2.91$ & $0.049$ & $26.0$ & $0.086$\\
         Type III & $8.19$ & $0.133$ & $12.9$ & $0.207$\\
         \hline\\
          \cline{2-3}\cline{4-5}HFB9& $\epsilon_{g9/2}$ & $\Gamma_{h11/2}$ & $\epsilon_{g9/2}$ & $\Gamma_{h11/2}$\\
        \hline Type I & $0.821$ & $0.050$  & $2.21$ & $0.077$\\
         Type II & $-2.99$ & $0.048$ & $21.4$ & $0.072$\\
         Type III & $8.18$ & $0.132$ & $12.8$ & $0.192$\\
         \hline Exp. & $5(12)$ & $0.059(18)$ & $33(12)$ & $0.116(17)$\\
    \end{tabular*}
    \label{tab:OpticalComparison}
\end{table}

\subsection{Density Profile Comparison}
\begin{figure}[tp]
    \centering
    \includegraphics[width=\linewidth]{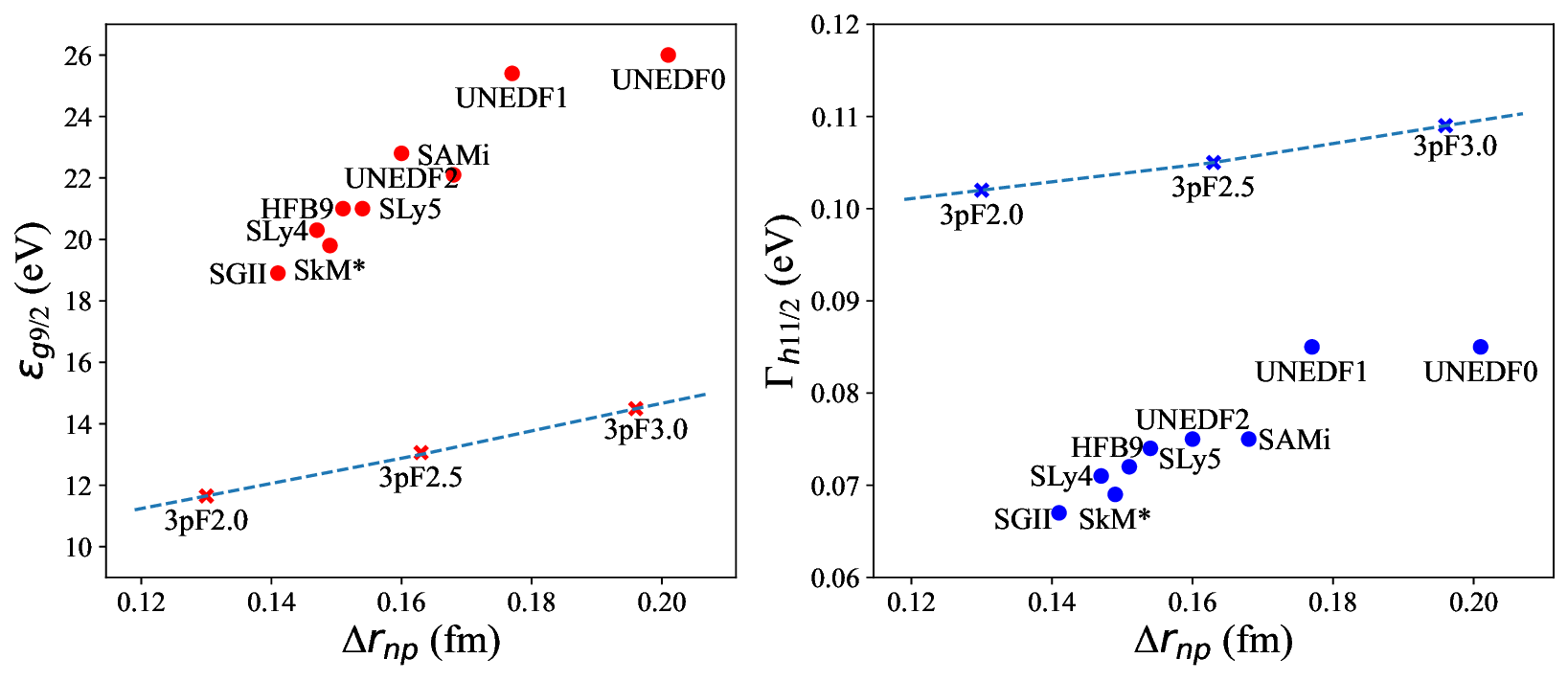}
    \caption{Scatter plots of the calculated values for $^{48}$Ca, with Type II parameter, as functions of the neutron skin thickness. Each point is labeled by used density profiles. The left and right panel demonstrates the result of shifts and widths, respectively.}
    \label{fig:drshift_typeII}
\end{figure}

In Table \ref{tab:densitycomparison}, we show the comparative results of the level shifts of $n=5$ and widths of $n=6$, with respect to each density profile, and with the Type II optical potential.
The second to fifth columns indicate the values for $^{40}$Ca, and right three ones are dedicated to $^{48}$Ca.
The neutron skin thickness $\Delta r_{np}$ is also presented next to the calculation results.
From this table one finds that, even if the magnitude of $\Delta r_{np}$ is comparable, both the strong shifts and widths can be significantly different, and particularly for $^{48}$Ca, the difference reaches a few tens percent among the different density profiles.
For visibility, in Fig.~\ref{fig:drshift_typeII}, calculated values are plotted as a function of neutron skin thickness, for $^{48}$Ca and with the Type II potential.
Each point is labeled by the name of its density profile.
From this figure it is found that the results of the 3pF calculations for both strong shift and level width gain almost linearly as neutron skin thickness swells up.
On the other hand, the calculation results with DFT-based density profiles seem to be positioned in a totally different sequence.
This figures also show that, when the 3pF profile is incorporated, while the shifts get smaller than those obtained with the density profiles based on DFT, the widths are significantly larger.
This indicates that quantitatively accurate predictions for the antiproton calcium spectrum will require a more detailed investigation that goes beyond the conventional 2pF or 3pF parametrizations and takes realistic density distributions into account.

\begin{table}[tbp]
    \centering
    \caption{A comparative result of the neutron skin thickness, strong shifts, and level widths, with respect to density profiles. For all the results Type II parameter is incorporated. The second to fourth, and fifth to seventh columns show the result of $^{40}$Ca and $^{48}$Ca, respectively. The neutron skin thickness is in a unit of fm, while the others are in eV.}
    \begin{tabular*}{0.7\columnwidth}{@{\extracolsep{\fill}}c|cccccc}
         &  \multicolumn{3}{c}{$^{40}$Ca} &  \multicolumn{3}{c}{$^{48}$Ca}\\
         \cline{2-4}\cline{5-7}& $\Delta r_{np}$ & $\varepsilon_{g9/2}$ & $\Gamma_{h11/2}$ & $\Delta r_{np}$ & $\varepsilon_{g9/2}$ & $\Gamma_{h11/2}$\\
         \hline\hline 3pF-2.0 & $2.32$ & $2.32$ & $0.136$ & 0.130 & $11.65$ & $0.102$\\
        3pF-2.5 & -- & -- & -- & $0.163$ & 13.06 & 0.105\\
        3pF-3.0 & -- & -- & -- & $0.196$ & 14.49 & 0.109\\
        \hline
        SLy4 & $-0.045$ & $-2.56$ & $0.047$ & $0.147$ & $20.29$ & $0.071$\\
        SLy5 & $-0.046$ & $-2.59$ & $0.047$ & $0.154$ & $20.99$ & $0.074$\\
        SkM* & $-0.047$ & $-2.95$ & $0.049$ & $0.149$ & $19.82$ & $0.069$\\
        SAMi & $-0.045$ & $-2.40$ & $0.042$ & $0.168$ & $22.13$ & $0.075$\\
        SGII & $-0.048$ & $-2.95$ & $0.045$ & $0.141$ & $18.92$ & $0.067$\\
        UNEDF0 & $-0.040$ & $-2.91$ & $0.049$ & $0.201$ & $25.98$ & $0.085$\\
        UNEDF1 & $-0.034$ & $-2.61$ & $0.045$ & $0.177$ & $25.40$ & $0.085$\\
        UNEDF2 & $-0.035$ & $-2.44$ & $0.043$ & $0.160$ & $22.81$ & $0.075$\\
        HFB9 & $-0.045$ & $-2.99$ & $0.048$ & $0.151$ & $21.05$ & $0.072$\\
         \hline
    \end{tabular*}
    \label{tab:densitycomparison}
\end{table}

\subsection{Nuclear State Effect}
Another interesting feature can be revealed from the Table~\ref{tab:densitycomparison}.
From strong shifts of \(^{40}\mathrm{Ca}\) with the Type II parameter, it is found that their signs are reversed from the 3pF model to other density distributions. 
It seems a little bit weird, because the optical potential of $^{40}$Ca is, as is shown in Fig.~\ref{fig:optical_typeII}, always attractive, which apparently does not contribute to repulsive shifts.
This feature can be explained by examining the ``nuclear'' states, namely the deeply bound states in the nuclear potential.
The effect of nuclear states on the atomic spectra has been already discussed in Ref.~\cite{gal1996}, where it has been pointed out that the existence of nuclear states brings about the dispersion behaviour in the strong shifts and widths.
Table~\ref{tab:nuclstate} lists the binding energies for several density distributions in \(^{40}\mathrm{Ca}\) with Type II parameters. 
As can be seen, the 3pF model has a deeply bound orbital with several MeV that does not appear in other density distributions.
This can be attributed to the lift down of the optical potential in the center region, which can be seen in Fig~\ref{fig:optical_typeII}.
This shallow bound states are inferred to contribute to level repulsion with the next higher atomic bound orbital, resulting in a slight shift of the binding energies.
The diagonalization method, provides all possible states without any prior bias, which enables us to discuss the energy level structures from multi-dimensional perspectives.
Additionally, this feature suggests that the shape of the optical potential can alter the value of the strong shift, due not only to its periphery structure but also to the behaviour in its center region.
This point should throw a stark light on the current framework of the $\bar{p}$-atom spectroscopy.


\begin{table}[htbp]
    \centering
    \caption{The binding energies for $j=9/2$ states of $^{40}$Ca, with Type II optical potential and several representative density profiles. All the energies are indicated in a unit of keV. "Nucl" means that the corresponding orbits are bound inside the nuclear potential, and "Atom" is outside.}
    \begin{tabular*}{0.5\columnwidth}{@{\extracolsep{\fill}}ccccc}
         & 3pF & SLy4 & SAMi & UNEDF0 \\
       \hline Nucl  & 3349 & -- & -- & --\\
        Atom  & 389.86 & 389.83 & 389.83 & 389.83\\
        \hline
    \end{tabular*}
    \label{tab:nuclstate}
\end{table}

\subsection{Spin-orbit Splitting}
\begin{figure}[htbp]
    \centering
    \includegraphics[width=0.5\columnwidth]{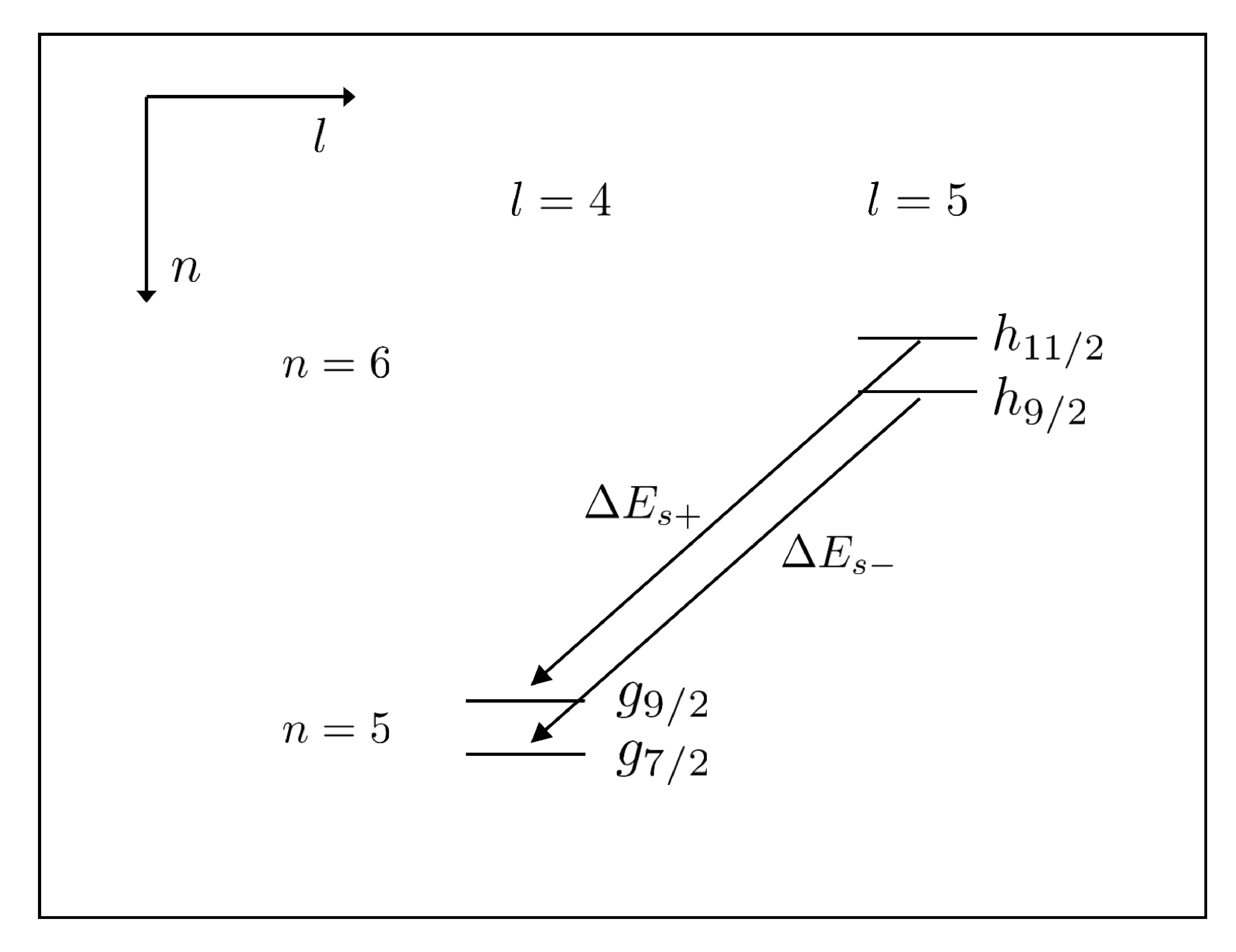}
    \caption{Level transition scheme in $\bar{p}^{40}$Ca and $\bar{p}^{48}$Ca.}
    \label{fig:energy-splitting}
\end{figure}

Figure \ref{fig:energy-splitting} shows the level transition scheme for $\bar{p}$-$^{40}$Ca and $\bar{p}$-$^{48}$Ca,
while the values of the transition energies from the yrast $n=6$ to $n=5$ states are demonstrated in Table \ref{tab:LS-splitting}.
Without considering the spin-flip transitions, the spin-orbit splittings should be observed as distinguished peaks with different spins.
These figure and table reveal that the magnitude of the spin-orbit splittings are around 235 eV for both nuclei,
which could be separated with the state-of-the-art high-resolution X-ray spectroscopy.
Furthermore, it is found that the transition energies are almost unchanged by choice of density profiles or optical potentials,
which indicates that the scale of spin-orbit splittings is not affected by detail of the potential shapes.
Additionally, we note that, when one considers the measurability of the spin-orbit splittings, the effect of the anomalous magnetic moment plays a crucial role.
This is because with the Pauli approximation the transition energies can be written as
\begin{equation}
    \Delta E_{nj} \approx (1+2\kappa)\frac{\mu c^2}{2} \frac{(Z\alpha)^4}{n^3 l(l+1)},
\end{equation}
which is explicitly proportional to $(1+2\kappa)$.
The anomalous moment of antiproton is approximately $\kappa\sim 1.8$, which leads to the gain of the spin-orbit splittings by a factor of $4.6$.
Without this effect the value of the splittings is around 50 eV, which should be obscured in the level widths of $n=5$ states.

\begin{table}[htbp]
    \centering
    \caption{The energy differences of the $\bar{p}\,6\to 5$ transitions. $\Delta E_{s+}$ denotes the transition from 6h$_{11/2}$ to 5g$_{9/2}$, while $\Delta E_{s-}$ is for 6h$_{9/2}$ to 5g$_{7/2}$ transition. The difference between $\Delta E_{s+}$ and $\Delta E_{s-}$ is also demonstrated as $\Delta E_{\pm}$. All the energies are exhibited in unit of keV. In the ``3pF'' column, for the density profile of $^{48}$Ca the 3pF-3.0 is adopted.}
    \begin{tabular*}{0.8\columnwidth}{@{\extracolsep{\fill}}cccc|ccc}
         & \multicolumn{3}{c}{$^{40}$Ca} & \multicolumn{3}{c}{$^{48}$Ca} \\
         \cline{2-4}\cline{5-7}& $\Delta E_{s+}$ & $\Delta E_{s-}$ & $\Delta E_{\pm}$ & $\Delta E_{s+}$ & $\Delta E_{s-}$ & $\Delta E_{\pm}$\\
         \hline\hline pure e.m.\\
         \hline 3pF & 119.093 & 119.327 &0.234  & 119.575 & 119.810 & 0.235 \\
         SLy4 & 119.094 & 119.328 &0.234  & 119.584 & 119.819 &0.235 \\
         UNEDF0 & 119.094 & 119.328 &0.234  & 119.584 & 119.819 &0.235 \\
         \hline w/ Type II opt.\\
         \hline 3pF & 119.102 & 119.336 & 0.234  & 119.562 & 119.797 & 0.235 \\
         SLy4 & 119.096 & 119.330 &0.234  & 119.562 & 119.797 &0.235 \\
         UNEDF0 & 119.096 & 119.331 &0.235  & 119.556 & 119.791 &0.235\\
         \hline\hline
    \end{tabular*}
    \label{tab:LS-splitting}
\end{table}

\section{DISCUSSION}
In the preceding sections we added an extra contribution to the isoscalar $b_0$ term and examined how it changes the calculated observables. Under the conventional global fitting framework the $b_1$ term has been regarded as zero, so the nonzero value found here is anomalous in that sense. The proton-antiproton scattering length is $b_{p\bar{p}} = 0.81 + 0.72i\,\fm$, and the value under discussion is roughly an order of magnitude larger. Even allowing for in-medium effects that might shift the effective scattering length, such a large number is not easy to accept. We therefore begin by seeing how much the p-wave $c_0$ term can be combined with $b_1$ to reduce the latter.

With that goal in mind we introduce the parameter sets Type IV and Type V listed in Table \ref{tab:pwaveparam}. Type IV retains the $c_0$ value of Type III, letting it shoulder part of the contribution to the spectral quantities. Type V extends the exploration to the real part of $c_0$ in hopes of lowering $b_1$ still further. Conversely, we also test whether enlarging $b_1$ alone can reproduce the experimental data: Type VI doubles the real part of $b_1$ in Type II, which may yield better results for the strong shift.

Table \ref{tab:additresult} presents the shifts and widths for the same representative density profiles used in Table \ref{tab:OpticalComparison}. To make the deviations from experiment clearer, entries lying within one standard deviation of the experimental value are underlined, and those within one half of a standard deviation are double-underlined.
The results show that, for every density profile, Types IV and V deliver quantitative improvements over the first three parameter sets. Nevertheless, the $^{48}$Ca observables are still not well reproduced: the strong shift stays near twenty and the widths fall outside the experimental standard deviation. For Type VI, the strong shift of $^{48}$Ca is well reproduced in magnitude for all density profiles, yet, as noted in Section 3.2, the dependence on the density distribution becomes much larger.
For example, the 3pF profile gives strong shift values not very different from those of Types IV and V, whereas for the UNEDF profile it gets far beyond the experimental value.
Additionally, the $^{40}$Ca strong shift gets too much repulsive, and the $^{48}$Ca width stays around 0.05-0.07 eV, far from the measured value.

These findings indicate that the $b_1$ and $c_0$ terms influence the spectral quantities in a complementary way, and that a quantitative explanation of the experimental data requires evaluating them from much more perspectives. Of course the present discussion is limited to calcium, and the experimental uncertainties for calcium are large, so the parameters cannot yet be pinned down uniquely. Even so, the complexity seen already for calcium suggests that future updates to the experimental data will permit more detailed analyses, and that similar studies on other nuclei will help clarify the relationship between the optical-model parameters and the properties they are meant to describe.

\begin{table}[tbp]
    \centering
        \caption{The optical potential parameters which are newly set for the discussion.}
    \begin{tabular*}{0.6\columnwidth}{@{\extracolsep{\fill}}cccc}
         & $b_0(\fm)$ & $b_1(\fm)$ & $c_0(\fm^{-3})$\\
        \hline\hline Type IV & $1.3+1.9i$ & $-4.0 + 0.0i$ & $0.0+1.2i$\\
        Type V & $1.3+1.9i$ & $-3.0+1.7i$ & $-1.5+0.5i$\\
        Type VI & $1.3+1.9i$ & $-16.0+1.7i$ & --\\
        \hline
    \end{tabular*}
    \label{tab:pwaveparam}
\end{table}


\begin{table}[t]
    \centering
    \caption{The strong shift for $n=5$ and level width for $n=6$ in a unit of eV, with respect to optical parameter sets. The second to the third column, the result of $^{40}$Ca is provided and the fourth to fifth column is dedicated to the counterpart of $^{48}$Ca with the 3pF, SLy4, UNEDF0 and HFB9 density. At the bottom line the experimental result\cite{hartmann2001a} is shown. The calculated values are underlined when they are within the standard error of from the experimental values, and double underlined when half standard error.}
    \begin{tabular*}{0.6\columnwidth}{@{\extracolsep{\fill}}ccccc}
         & \multicolumn{2}{c}{$^{40}$Ca} & \multicolumn{2}{c}{$^{48}$Ca} \\
         \cline{2-3}\cline{4-5} 3pF & $\epsilon_{g9/2}$ & $\Gamma_{h11/2}$ & $\epsilon_{g9/2}$ & $\Gamma_{h11/2}$\\
         \hline Type IV & $\uuline{7.73}$ & $0.237$  & $14.2$ & $0.176$\\
         Type V & $\uline{16.1}$ & $0.142$ & $19.5$ & $\uline{0.104}$\\
         Type VI & $\uuline{2.32}$ & $0.136$ & $\uline{24.9}$ & $0.075$\\
         \hline\\
         \cline{2-3}\cline{4-5}SLy4& $\epsilon_{g9/2}$ & $\Gamma_{h11/2}$ & $\epsilon_{g9/2}$ & $\Gamma_{h11/2}$\\
        \hline Type IV & $\uuline{7.10}$ & $0.132$  & $18.1$ & $0.165$\\
         Type V & $\uline{16.0}$ & $\uuline{0.059}$ & $\uline{25.0}$ & $0.081$\\
         Type VI & $\uline{-5.90}$ & $\uuline{0.052}$ & $\uuline{33.6}$ & $0.055$\\
         \hline\\
         \cline{2-3}\cline{4-5} UNEDF0& $\epsilon_{g9/2}$ & $\Gamma_{h11/2}$ & $\epsilon_{g9/2}$ & $\Gamma_{h11/2}$\\
        \hline Type IV & $\uuline{6.90}$ & $0.136$  & $20.7$ & $0.176$\\
         Type V & $\uline{16.0}$ & $\uuline{0.062}$ & $\uline{26.4}$ & $0.092$\\
         Type VI & $\uline{-6.49}$ & $\uuline{0.053}$ & $\uline{42.3}$ & $0.067$\\
         \hline\\
          \cline{2-3}\cline{4-5}HFB9& $\epsilon_{g9/2}$ & $\Gamma_{h11/2}$ & $\epsilon_{g9/2}$ & $\Gamma_{h11/2}$\\
        \hline Type IV & $\uuline{6.87}$ & $0.135$  & $18.4$ & $0.165$\\
         Type V & $\uline{16.0}$ & $\uuline{0.061}$ & $\uline{25.0}$ & $0.082$\\
         Type VI & $\uuline{-6.66}$ & $\uuline{0.053}$ & $\uuline{34.8}$ & $0.056$\\
         \hline Exp. & $5(12)$ & $0.059(18)$ & $33(12)$ & $0.116(17)$\\
    \end{tabular*}
    \label{tab:additresult}
\end{table}

\section{SUMMARY}
In this work, a comprehensive study in the $\bar{p}$-Ca atom spectroscopy has been conducted based on the Dirac equation with optical potentials which include the isovector or p-wave components as well as with realistic density profiles via the nuclear DFT calculations incorporated.
A comparison of the calculated strong shifts and level widths obtained with several optical-potential parameter sets shows that reproducing the systematic differences between the experimental values for $^{40}$Ca and $^{48}$Ca requires the inclusion of various additional terms, such as the p-wave contribution and the isovector term that had previously been considered unnecessary.  Once the isovector term is introduced, the calculated results are found to change significantly—by several tens of percent—depending on the choice of density profile.  These variations arise not only from quantitative changes in the shape of the potential but also from modifications to the concrete structure of the deeply bound levels.
Taken together, these findings argue that predictions of antiprotonic atomic spectra based on the optical model must adopt a more comprehensive perspective, and they offer a fresh viewpoint for future research in this field.  In practice, investigators should go beyond statistical parameter fitting and carry out systematic studies for each individual nucleus, analyzing in detail the relationship between every parameter and the resulting observables, including cases that at first sight appear anomalous.  As experimental techniques advance and more precise data become available, such systematic analyses will yield deeper insight.  Rapid progress in studies of nuclear density distributions and neutron-skin thickness also promises to supply additional information that could shed new light on the problem.
By pursuing complementary and integrated investigations of this kind, we may finally move closer to a full understanding of nuclear structure and of the underlying nature of antiproton-nucleon, and more generally baryon-baryon, interactions.

\section*{Acknowledgement}
We would extend gratitude to Hiroyuki Fujioka for valuable discussions and providing knowledge from the experimental field.
We would also thank Tomoya Naito for offering calculation data of density distributions.
This work partly made use of computational resources of the Yukawa-21 supercomputer at Yukawa Institute for Theoretical Physics (YITP), Kyoto University.
One of the authors (K.Y.) would like to acknowledge the financial support from the JSPS Research Fellow, Grant No. 24KJ1110.
The work of S.Y. was supported by JST SPRING No. JPMJSP2180.
The work of D.J. was partly supported by Grants-in-Aid for Scientific Research from JSPS (21K03530, 22H04917, 23K03427).
The author J.Y. acknowledges the financial support by Grants-in-Aid for scientific Research from JSPS (22K03607).

\appendix 




\section{Full Set of Calculation Results}\label{Sec:FullResult}
Table \ref{tab:fullresult_g9/2} to \ref{tab:fullresult_h9/2} show the full set of calculation results for strong shifts $\varepsilon$ and level widths $\Gamma$ of 5g$_{9/2}$, 5g$_{7/2}$, 6h$_{11/2}$ and 6h$_{9/2}$ states.
From these tables, it is clear that both the strong shifts and level widths do not change between the different spin states,
which is why in the body the spin dependence of the spectroscopy has not been particularly discussed.
In addition, the level widths for $n=5$ states are around 30-40 eV, which is more or less consistent with a weighted average of 35(6) eV for $^{40,42,43,44,48}$Ca~\cite{hartmann2001a}.


\begin{table}[htbp]
    \centering
    \caption{The full calculation results of level shifts and widths of the $g_{9/2}$ state, with respect to the optical potential parameters and density profiles. The second to the fifth columns, the shift and width for $^{40}$Ca, the counterparts for $^{48}$Ca are provided, with the Type I potential. The sixth to tenth, eleventh to fourteenth columns are dedicated for Type II and Type III potential in the same order. The result of 3pF density for $^{40}$Ca is exhibited in "3pF-2.0" line.}
    \scriptsize
    \begin{tabular*}{1\linewidth}{@{\extracolsep{\fill}}ccccccccccccc}
         & \multicolumn{4}{c}{Type I} & \multicolumn{4}{c}{Type II} & \multicolumn{4}{c}{Type III} \\
         \cline{2-5}\cline{6-9}\cline{10-13}\cline{2-5}\cline{6-9}\cline{10-13}& \multicolumn{2}{c}{$^{40}$Ca} & \multicolumn{2}{c}{$^{48}$Ca} & \multicolumn{2}{c}{$^{40}$Ca} & \multicolumn{2}{c}{$^{48}$Ca} & \multicolumn{2}{c}{$^{40}$Ca} & \multicolumn{2}{c}{$^{48}$Ca}\\
        $g_{9/2}$& $\varepsilon(\mathrm{eV})$ & $\Gamma(\mathrm{eV})$ & $\varepsilon(\mathrm{eV})$ & $\Gamma(\mathrm{eV})$ & $\varepsilon(\mathrm{eV})$ & $\Gamma(\mathrm{eV})$ & $\varepsilon(\mathrm{eV})$ & $\Gamma(\mathrm{eV})$ & $\varepsilon(\mathrm{eV})$ & $\Gamma(\mathrm{eV})$ & $\varepsilon(\mathrm{eV})$ & $\Gamma(\mathrm{eV})$\\
         \hline    3pF-2.0 & 2.32 & 44.58 & 3.67 & 36.49 & 2.32  & 44.58 & 11.81 & 25.85 & 7,73  & 64.6  & 9.99 & 53.61 \\
    3pF-2.5 & -- & -- & 3.83 & 37.52 & --  & -- & 13.02 & 25.98 & --  & --  & 10.34 & 54.99 \\
    3pF-3.0 & -- & -- & 3.99 & 38.62 & --  & -- & 14.2  & 26.16 & -- & --  & 10.71 & 56.45 \\
    SLy4   & 0.94 & 20.86 & 2.32 & 29.11 & $-2.56$ & 21.99 & 20.29 & 19.77 & 8.26  & 41.54 & 12.88 & 55.59 \\
    SLy5   & 0.92 & 20.74 & 2.25 & 29.6  & $-2.59$ & 21.88 & 20.99 & 20.19 & 8.21  & 41.33 & 12.86 & 56.24 \\
    SkM*   & 0.85 & 21.43 & 2.27 & 28.86 & $-2.95$ & 22.59 & 19.82 & 19.39 & 8.24  & 42.33 & 12.76 & 55.18 \\
    SAMi   & 0.91 & 19.27 & 2.18 & 29.56 & $-2.4$  & 20.33 & 22.13 & 20.09 & 7.94  & 39.11 & 12.76 & 56.15 \\
    SGII   & 0.74 & 20.03 & 2.07 & 27.74 & $-2.95$ & 21.12 & 18.92 & 18.88 & 7.84  & 40.11 & 12.33 & 53.43 \\
    UNEDF0 & 0.78 & 21.49 & 2.0  & 31.91 & $-2.91$ & 22.31 & 25.98 & 21.69 & 8.19  & 42.42 & 12.9  & 59.36 \\
    UNEDF1 & 0.67 & 20.4  & 1.75 & 30.79 & $-2.61$ & 20.93 & 25.4  & 21.5  & 7.85  & 40.65 & 12.39 & 57.55 \\
    UNEDF2 & 0.71 & 19.56 & 1.88 & 28.7  & $-2.44$ & 20.13 & 22.81 & 19.87 & 7.77  & 39.43 & 12.26 & 54.7 \\
    HFB9   & 0.82 & 21.28 & 2.21 & 29.24 & $-2.99$ & 22.27 & 21.05 & 19.87 & 8.18  & 42.09 & 12.75 & 55.69 \\
         \hline
    \end{tabular*}
    \label{tab:fullresult_g9/2}
\end{table}

\begin{table}[htbp]
    \centering
    \caption{The same with Table \ref{tab:fullresult_g9/2}, but for $g_{7/2}$ state.}
    \scriptsize
    \begin{tabular*}{1\columnwidth}{@{\extracolsep{\fill}}ccccccccccccc}
         & \multicolumn{4}{c}{Type I} & \multicolumn{4}{c}{Type II} & \multicolumn{4}{c}{Type III} \\
         \cline{2-5}\cline{6-9}\cline{10-13}\cline{2-5}\cline{6-9}\cline{10-13}& \multicolumn{2}{c}{$^{40}$Ca} & \multicolumn{2}{c}{$^{48}$Ca} & \multicolumn{2}{c}{$^{40}$Ca} & \multicolumn{2}{c}{$^{48}$Ca} & \multicolumn{2}{c}{$^{40}$Ca} & \multicolumn{2}{c}{$^{48}$Ca}\\
        $g_{7/2}$& $\varepsilon(\mathrm{eV})$ & $\Gamma(\mathrm{eV})$ & $\varepsilon(\mathrm{eV})$ & $\Gamma(\mathrm{eV})$ & $\varepsilon(\mathrm{eV})$ & $\Gamma(\mathrm{eV})$ & $\varepsilon(\mathrm{eV})$ & $\Gamma(\mathrm{eV})$ & $\varepsilon(\mathrm{eV})$ & $\Gamma(\mathrm{eV})$ & $\varepsilon(\mathrm{eV})$ & $\Gamma(\mathrm{eV})$\\
        \hline 3pF-2.0 & 2.31 & 45.21  & 3.67 & 36.99  & 2.31  & 45.21  & 11.94 & 26.21  & 8.75  & 65.56  & 11.93 & 54.39 \\
    3pF-2.5 & --   & --    & 3.83 & 38.04 & --    & --    & 13.16 & 26.34  & --    & --    & 12.31 & 55.79 \\
    3pF-3.0 & --   & --    & 4.0  & 39.15 & --    & --    & 14.36 & 26.53  & --    & --    & 12.7  & 57.27 \\
    SLy4   & 0.9  & 21.1  & 2.29 & 29.48 & $-2.65$ & 22.25 & 20.55 & 20.04 & 8.33  & 42.1  & 13.01 & 56.38 \\
    SLy5   & 0.88 & 20.98 & 2.22 & 29.99 & $-2.69$ & 22.13 & 21.26 & 20.47 & 8.27  & 41.89 & 12.98 & 57.04 \\
    SkM*   & 0.81 & 21.69 & 2.25 & 29.23 & $-3.05$ & 22.86 & 20.07 & 19.65 & 8.3   & 42.91 & 12.89 & 55.96 \\
    SAMi   & 0.88 & 19.5  & 2.15 & 29.95 & $-2.48$ & 20.56 & 22.42 & 20.36 & 8.0   & 39.64 & 12.89 & 56.94 \\
    SGII   & 0.7  & 20.27 & 2.05 & 28.09 & $-3.05$ & 21.37 & 19.16 & 19.14 & 7.9   & 40.66 & 12.45 & 54.19 \\
    UNEDF0 & 0.75 & 21.75 & 1.97 & 32.34 & $-3.01$ & 22.58 & 26.33 & 21.99 & 8.25  & 43.0  & 13.03 & 60.21 \\
    UNEDF1 & 0.64 & 20.64 & 1.72 & 31.2  & $-2.71$ & 21.18 & 25.75 & 21.79 & 7.91  & 41.21 & 12.52 & 58.38 \\
    UNEDF2 & 0.68 & 19.79 & 1.85 & 29.07 & $-2.53$ & 20.36 & 23.11 & 20.14 & 7.83  & 39.97 & 12.38 & 55.47 \\
    HFB9   & 0.78 & 21.53 & 2.18 & 29.61 & $-3.09$ & 22.54 & 21.32 & 20.14 & 8.24  & 42.66 & 12.88 & 56.48 \\
        \hline        
    \end{tabular*}
    \label{tab:fullresult_g7/2}
\end{table}

\begin{table}[htbp]
    \centering
    \caption{The same with Table \ref{tab:fullresult_g9/2}, but for $h_{11/2}$ state.}
    \scriptsize
    \begin{tabular*}{\columnwidth}{@{\extracolsep{\fill}}ccccccccccccc}
         & \multicolumn{4}{c}{Type I} & \multicolumn{4}{c}{Type II} & \multicolumn{4}{c}{Type III} \\
         \cline{2-5}\cline{6-9}\cline{10-13}\cline{2-5}\cline{6-9}\cline{10-13}& \multicolumn{2}{c}{$^{40}$Ca} & \multicolumn{2}{c}{$^{48}$Ca} & \multicolumn{2}{c}{$^{40}$Ca} & \multicolumn{2}{c}{$^{48}$Ca} & \multicolumn{2}{c}{$^{40}$Ca} & \multicolumn{2}{c}{$^{48}$Ca}\\
        $h_{11/2}$& $\varepsilon(\mathrm{eV})$ & $\Gamma(\mathrm{eV})$ & $\varepsilon(\mathrm{eV})$ & $\Gamma(\mathrm{eV})$ & $\varepsilon(\mathrm{eV})$ & $\Gamma(\mathrm{eV})$ & $\varepsilon(\mathrm{eV})$ & $\Gamma(\mathrm{eV})$ & $\varepsilon(\mathrm{eV})$ & $\Gamma(\mathrm{eV})$ & $\varepsilon(\mathrm{eV})$ & $\Gamma(\mathrm{eV})$\\
        \hline 3pF-2.0 & 0.0  & 0.136 & 0.003 & 0.102 & 0.0   & 0.136 & 0.027 & 0.088  & 0.012 & 0.238 & 0.019 & 0.185 \\
    3pF-2.5 & --   & --    & 0.003 & 0.105 & --    & --    & 0.031 & 0.090 & --    & --    & 0.02  & 0.191 \\
    3pF-3.0 & --   & --    & 0.003 & 0.109 & --    & --    & 0.034 & 0.093 & --    & --    & 0.021 & 0.198 \\
    SLy4   & $-0.006$ & 0.049  & $-0.007$ & 0.076  & $-0.019$ & 0.047  & 0.066  & 0.071 & 0.008 & 0.129 & 0.016 & 0.191 \\
    SLy5   & $-0.006$ & 0.048  & $-0.007$ & 0.078  & $-0.019$ & 0.047  & 0.07   & 0.074 & 0.008 & 0.128 & 0.016 & 0.194 \\
    SkM*   & $-0.006$ & 0.051  & $-0.007$ & 0.075  & $-0.021$ & 0.049  & 0.063  & 0.069 & 0.008 & 0.132 & 0.016 & 0.189 \\
    SAMi   & $-0.005$ & 0.044  & $-0.007$ & 0.078  & $-0.018$ & 0.042  & 0.075  & 0.075 & 0.008 & 0.119 & 0.016 & 0.194 \\
    SGII   & $-0.006$ & 0.047  & $-0.007$ & 0.072  & $-0.02$  & 0.045  & 0.06   & 0.067 & 0.007 & 0.124 & 0.015 & 0.182 \\
    UNEDF0 & $-0.007$ & 0.051  & $-0.009$ & 0.087  & $-0.022$ & 0.049  & 0.092  & 0.085 & 0.007 & 0.133 & 0.015 & 0.21 \\
    UNEDF1 & $-0.006$ & 0.048  & $-0.009$ & 0.083  & $-0.02$  & 0.045  & 0.09   & 0.085 & 0.007 & 0.126 & 0.014 & 0.202 \\
    UNEDF2 & $-0.006$ & 0.046  & $-0.008$ & 0.076  & $-0.018$ & 0.043  & 0.078  & 0.075 & 0.007 & 0.121 & 0.015 & 0.189 \\
    HFB9   & $-0.006$ & 0.051  & $-0.007$ & 0.077  & $-0.022$ & 0.048  & 0.069  & 0.072 & 0.007 & 0.132 & 0.016 & 0.192 \\
        \hline        
    \end{tabular*}
    \label{tab:fullresult_h11/2}
\end{table}

\begin{table}[htbp]
    \centering
    \caption{The same with Table \ref{tab:fullresult_g9/2}, but for $h_{9/2}$ state.}
    \scriptsize
    \begin{tabular*}{\columnwidth}{@{\extracolsep{\fill}}ccccccccccccc}
         & \multicolumn{4}{c}{Type I} & \multicolumn{4}{c}{Type II} & \multicolumn{4}{c}{Type III} \\
         \cline{2-5}\cline{6-9}\cline{10-13}\cline{2-5}\cline{6-9}\cline{10-13}& \multicolumn{2}{c}{$^{40}$Ca} & \multicolumn{2}{c}{$^{48}$Ca} & \multicolumn{2}{c}{$^{40}$Ca} & \multicolumn{2}{c}{$^{48}$Ca} & \multicolumn{2}{c}{$^{40}$Ca} & \multicolumn{2}{c}{$^{48}$Ca}\\
        $h_{9/2}$& $\varepsilon(\mathrm{eV})$ & $\Gamma(\mathrm{eV})$ & $\varepsilon(\mathrm{eV})$ & $\Gamma(\mathrm{eV})$ & $\varepsilon(\mathrm{eV})$ & $\Gamma(\mathrm{eV})$ & $\varepsilon(\mathrm{eV})$ & $\Gamma(\mathrm{eV})$ & $\varepsilon(\mathrm{eV})$ & $\Gamma(\mathrm{eV})$ & $\varepsilon(\mathrm{eV})$ & $\Gamma(\mathrm{eV})$\\
        \hline 3pF-2.0 & 0.0   & 0.136 & 0.003 & 0.101  & 0.0   & 0.136 & 0.027 & 0.088  & 0.012 & 0.239 & 0.019 & 0.185 \\
    3pF-2.5 & --    & --    & 0.003 & 0.105 & --    & --    & 0.031 & 0.091 & --    & --    & 0.02  & 0.182 \\
    3pF-3.0 & --    & --    & 0.003 & 0.109 & --    & --    & 0.034 & 0.093 & --    & --    & 0.021 & 0.198 \\
    SLy4   & $-0.006$ & 0.048 & $-0.007$ & 0.076 & $-0.02$ & 0.047 & 0.066 & 0.071 & 0.008 & 0.129 & 0.016 & 0.191 \\
    SLy5   & $-0.006$ & 0.048 & $-0.007$ & 0.078 & $-0.02$ & 0.046 & 0.07  & 0.074 & 0.008 & 0.128 & 0.016 & 0.195 \\
    SkM*   & $-0.006$ & 0.051 & $-0.007$ & 0.075 & $-0.022$ & 0.049 & 0.063 & 0.069 & 0.007 & 0.132 & 0.016 & 0.189 \\
    SAMi   & $-0.005$ & 0.044 & $-0.008$ & 0.078 & $-0.018$ & 0.042 & 0.075 & 0.075 & 0.008 & 0.119 & 0.015 & 0.195 \\
    SGII   & $-0.006$ & 0.046 & $-0.007$ & 0.071 & $-0.021$ & 0.044 & 0.06  & 0.067 & 0.007 & 0.124 & 0.015 & 0.182 \\
    UNEDF0 & $-0.007$ & 0.051 & $-0.009$ & 0.087 & $-0.022$ & 0.048 & 0.092 & 0.086 & 0.007 & 0.133 & 0.014 & 0.21  \\
    UNEDF1 & $-0.006$ & 0.048 & $-0.009$ & 0.083 & $-0.02$  & 0.045 & 0.091 & 0.085 & 0.007 & 0.126 & 0.014 & 0.203 \\
    UNEDF2 & $-0.006$ & 0.045 & $-0.008$ & 0.075 & $-0.019$ & 0.043 & 0.078 & 0.075 & 0.007 & 0.121 & 0.014 & 0.189 \\
    HFB9   & $-0.006$ & 0.05  & $-0.007$ & 0.076 & $-0.022$ & 0.048 & 0.069 & 0.072 & 0.007 & 0.132 & 0.016 & 0.192 \\
        \hline        
    \end{tabular*}
    \label{tab:fullresult_h9/2}
\end{table}

\bibliographystyle{ptephy}
\bibliography{antiproton}

\end{document}